\documentclass{icrc}

\usepackage{times}
\usepackage[dvips]{graphicx} 
\usepackage{amsmath}
\begin{document}

\title{Mass composition of cosmic rays in anomalous diffusion model: comparison
with experiment}
\author[1]{A.A. Lagutin}
\author[1]{D.V. Strelnikov}
\author[1]{A.G. Tyumentsev}
\affil[1]{Altai State University. Barnaul 656099, Russia}

\correspondence{Lagutin (lagutin@theory.dcn-asu.ru)}

\firstpage{1}
\pubyear{2001}


\maketitle

\begin{abstract}
We calculate the energy spectra and mass composition of cosmic
rays in energy region ($1\div 10^8$)GeV/particle under the
assumption that cosmic rays propagation in the Galaxy is described
by anomalous diffusion equation. Our results and comparisons with
experimental data are presented.
\end{abstract}

\section{Introduction}
The steepening of the all-particle spectrum around $3\cdot 10^{15}
eV$ (the ``knee'') discovered in 1958 \citep{KandK:1958} has been
the subject of numerous speculations on the propagation and
acceleration mechanisms of galactic cosmic rays (see, for example,
the reviews by \citet{Erlykin}, \citet{KandK:1995},
\citet{Ptuskin:1997}).  The usually accepted picture of cosmic ray
propagation in the interstellar medium  is a normal diffusion,
which can be described by equation for concentration
\citep{Ginzburg and Syr:1964, Berezinsky:1990}
\begin{equation}\label{u1}
  \frac{\partial N}{\partial t}=D(E)\Delta
  N(\vec{r},t,E)+S(\vec{r},t,E),
\end{equation}
The ``knee'' is not an intrinsic property of this model under the
natural physical conditions $D(E)\sim E^\delta$, $S(E)\sim
E^{-p}$. Because of these reasons discussions have mainly been
directed to a search and a justification of breakdown mechanisms
either of the diffusivity $D(E)$ or of the source density $S(E)$.
However, in spite of considerable theoretical and experimental
efforts, a model, which can explain
\begin{enumerate}
\item
different values of spectral exponent of protons and other nuclei,
\item
mass composition variations at $E\sim 10^2\div 10^5$
\\ GeV/nucleon,
\item
the steepening of the all-particle spectrum
\end{enumerate}
was not developed.

Recently, in our papers
\citep{LagNikUch:2000,LagNikUch:2001,LNUIzvRAN:2001} new view of
the ``knee'' problem was presented. It has been shown that the
``knee'' in the primary cosmic rays spectrum is due to fractal
structure of the interstellar medium, that is another regime of
particles diffusion in the Galaxy.

In this paper we consider the propagation of galactic cosmic rays
in the fractal interstellar medium taking into account that a
particle can spend long time in a trap. We demonstrate the main
results of this new model in a wide energy region $E\sim (1\div
10^8)$ Gev/nucleon.
\section{Model}

Based on the results \citep{LU:this conf.} the cosmic ray
propagation in the fractal interstellar medium is described by
fractional diffusion equation. Without energy losses and nuclear
interactions, the equation for concentration of the cosmic rays
with energy $E$ generated by sources $S(\vec{r},t,E)$ has the form
\begin{equation}\label{fracdifeq}
  \frac{\partial N}{\partial
  t}=-D(E,\alpha,\beta)D_{0+}^{1-\beta}(-\Delta)^{\alpha/2}N(\vec{r},t,E)+S(\vec{r},t,E)
\end{equation}
where $D(E,\alpha,\beta)$ is the anomalous diffusivity, $\alpha$
and $\beta$ are determined by the fractal structure of the medium
and by the trapping mechanism, correspondingly (see \citet{LU:this
conf.}). $D_{0+}^\mu$ denotes the Riemann-Liouville fractional
derivative, $(-\Delta)^{\alpha/2}$
--- the fractional Laplacian \citep{samko}.

In the case of point impulse source with inverse power spectrum,
relating to supernova bursts
\[
 S(\vec{r},t,E)=S_0E^{-p}\delta(\vec{r})\Theta(T-t)\Theta(t),
\Theta(\tau)=\begin{cases} 1,&\tau>0,\\ 0,&\tau<0, \end{cases}
\]
the solution of equation (\ref{fracdifeq}), found in
\citep{LagNikUch:2001} is of the form
\begin{multline}\label{N(r,t,E}
N(\vec{r},t,E)=\frac{S'_0E^{-p}}{D(E,\alpha,\beta)^{3/\alpha}}\int\limits_{\max[0,t-T]}^{t}\tau^{-3\beta/\alpha}\\
\times\Psi_3^{(\alpha,\beta)}\Bigl({|\vec{r}|(D(E,\alpha,\beta)\tau^\beta)^{-1/\alpha}}\Bigr)d\tau,
\end{multline}
where the scaling function  $\Psi_3^{(\alpha,\beta)}(r)$,
\begin{equation}\label{psi3}
  \Psi_3^{(\alpha,\beta)}(r)=\int\limits_0^\infty{q_3^{(\alpha)}({r\tau^\beta})q_1^{(\beta,1)}(\tau)\tau^{3\beta/\alpha}d\tau},
\end{equation}
is determined by three-dimensional spherically-symmetrical stable
distribution $q_3^{(\alpha)}(r)\ (\alpha\leq 2)$ and one-sided
stable distribution $q_1^{(\beta,1)}(t)$ with characteristic
exponent $\beta$ \citep{Znanie:1983, Chance:1999}.

The anomalous diffusivity $D(E,\alpha,\beta)$ is determined by the
constants $A$ and $B$ in the asymptotic behaviour for ``L\'{e}vy
flights'' $(A)$ and ``L\'{e}vy waiting time'' $(B)$ distributions:
\[
D(E,\alpha,\beta)\propto A(E,\alpha)/B(E,\beta).
\]
Taking into account that both the free path and the probability to
stay in trap during the time interval $t$ for particle with charge
$Z$ and mass number $A$ depend on particle magnetic rigidity $R$,
we accept $D=(\upsilon/c)D_0(\alpha,\beta) R^\delta$.

Using the representation $N=N_0E^{-\eta}$ and the property
$d\Psi_m^{(\alpha,\beta)}/dr=-2\pi r\Psi_{m+2}^{(\alpha,\beta)}$
of the scaling function \citep{Chance:1999}, one can easy find the
spectral exponent $\eta$ for observed particles:
\begin{equation}\label{eta}
  \eta=p+\frac{\delta}{\alpha}\Xi,
\end{equation}
where
\begin{multline}\label{Xi}
  \Xi=3-\frac{2\pi r^2}{D(E,\alpha,\beta)^{2/\alpha}} \\
  \times\frac{\int\limits_{\max[0,t-T]}^{t}\tau^{-5\beta/\alpha}
\Psi_5^{(\alpha,\beta)}\Bigl({|\vec{r}|(D(E,\alpha,\beta)\tau^\beta)^{-1/\alpha}}\Bigr)}{\int\limits_{\max[0,t-T]}^{t}\tau^{-3\beta/\alpha}
\Psi_3^{(\alpha,\beta)}\Bigl({|\vec{r}|(D(E,\alpha,\beta)\tau^\beta)^{-1/\alpha}}\Bigr)}.
\end{multline}
Let $E_0$ be a solution of the equation $\Xi(E)=0$. One can see
from (\ref{eta})-(\ref{Xi}) that at $E=E_0$ the spectral exponent
for observed particles $\eta$ is equal to spectral exponent for
particles generated by the source: $\eta(E_0)=p$. Since  the
exponent $\eta_{E\ll E_0}=p-\delta$ is less than p at $E\ll E_0$,
but the exponent $\eta_{E\gg E_0}=p+\delta/\beta>p$ at $E\gg E_0$,
$E_0$ can be called the ``knee'' energy.

From experimental values of $\eta_{E\ll E_0}$ and $\eta_{E\gg
E_0}$ one can derive the main parameters of the model ($p,\delta$)
versus the spectral exponent ($\beta$) of ``the L\'{e}vy waiting
time'':
\[
\delta=(\eta_{E\gg E_0}-\eta_{E\ll E_0})\frac{\beta}{1+\beta},\
p=\eta_{E\ll E_0}+\delta.
\]

To evaluate the parameter $\beta$ we have used the results
presented in the paper \citep{Cadavid:1999}, where an anomalous
diffusion of solar magnetic elements have been investigated. The
authors have shown that the trapping time distribution
asymtotically takes the form of a L\'{e}vy distribution with
spectral exponent $\beta\approx 0.8$.

Assuming that a trapping mechanism is characterized by a kind of
self-similarity, one can expect the some value for $\beta$ in the
scales under the consideration . By this reason the value
$\beta=0.8$ is used in our calculations.

Thus, taking $\eta_{E\ll E_0}\sim 2.63$ and $\eta_{E\gg E_0}\sim
3.24$ we finally obtain $p\approx 2.9$, $\delta\approx 0.27$.

To evaluate the next important parameter --- anomalous diffusivity
$D_0(\alpha,\beta)$, we have used the experimental data on the
particle anisotropy in the the energy region $10^3\div 10^4$
GeV/particle in the framework of the scheme proposed by
\citet{Osborn:1976} and \citet{Dorman:1985}. For example, we find
$D_0\approx (1\div 4)\cdot 10^{-3}\ pc^{1.7}y^{-0.8}$ in the case
$\alpha =1.7,\ \beta=0.8$ for near sources Monogem ($r\sim 300\
pc,\ t\sim 10^5\ y$), Geminga ($r\sim 300\ pc,\ t\sim 3\cdot 10^5\
y$), Loop I-IV ($r\sim (100\div 200)\ pc,\ t\sim (2\div 4)\cdot
10^5\ y$).

In the model under consideration only one parameter $\alpha\
(1<\alpha< 2)$ connected with the fractal structure of the
interstellar medium is found by fit. Extensive calculations of
cosmic-ray spectra show that the best fit of experimental data may
be get at $\alpha\approx 1.7$.
\begin{figure*}[pb]
\includegraphics[width=8.3cm]{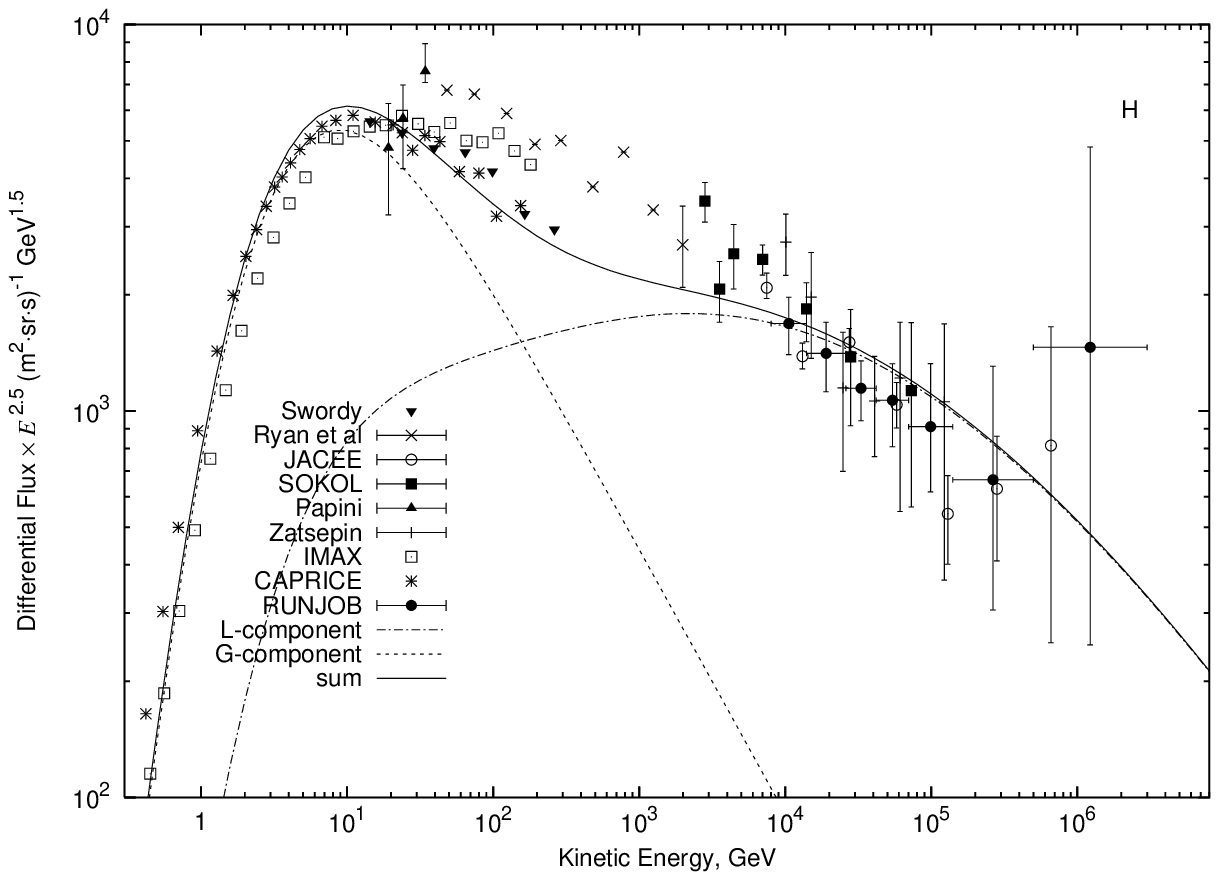}
\includegraphics[width=8.3cm]{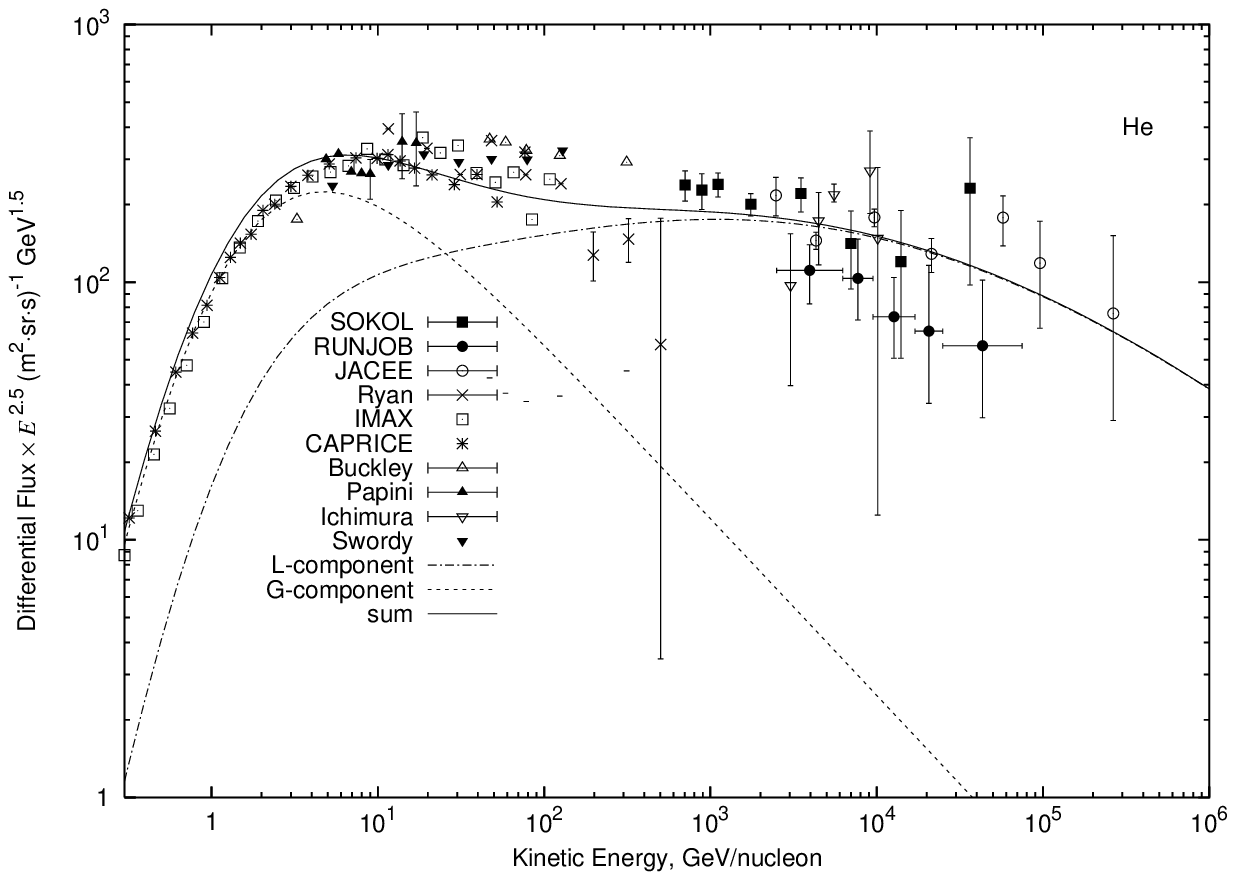}
\includegraphics[width=8.3cm]{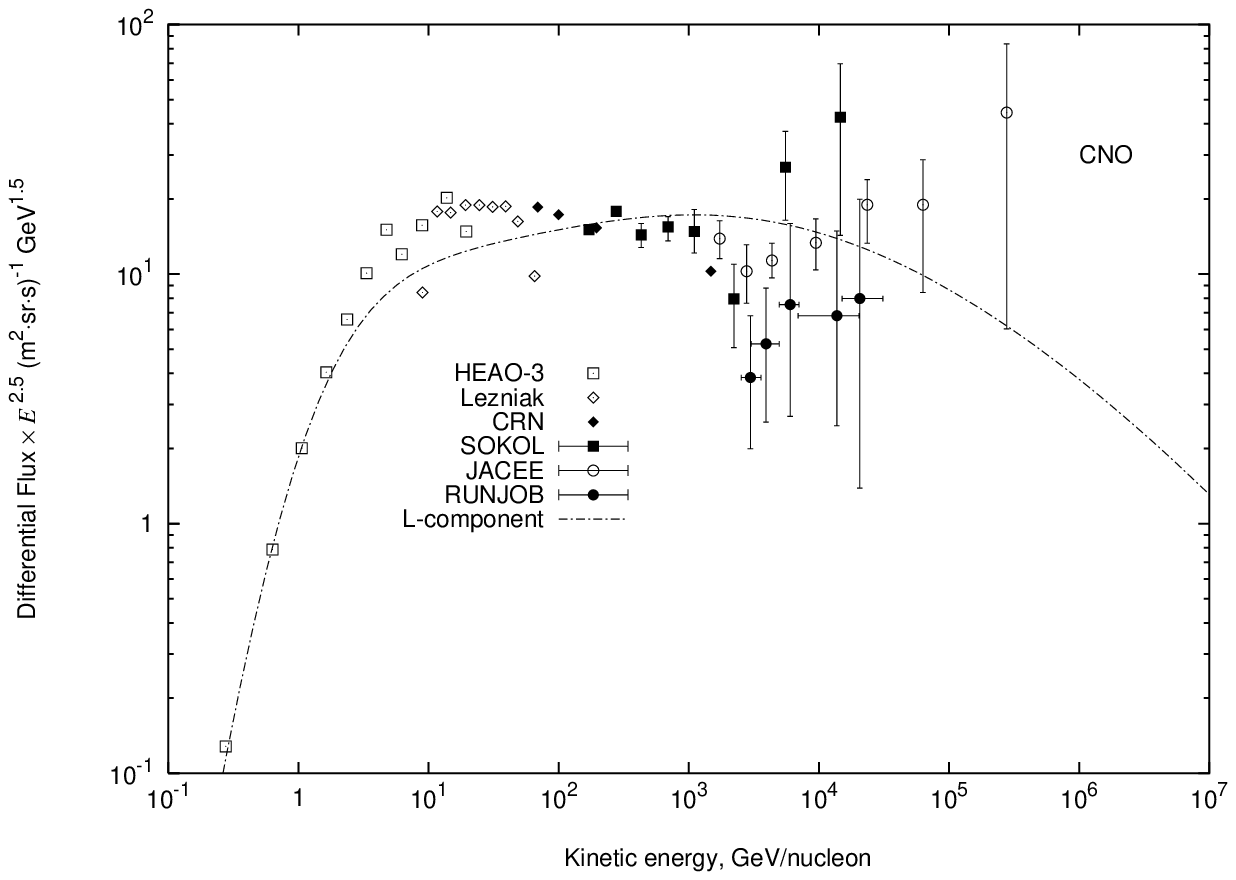}
\includegraphics[width=8.3cm]{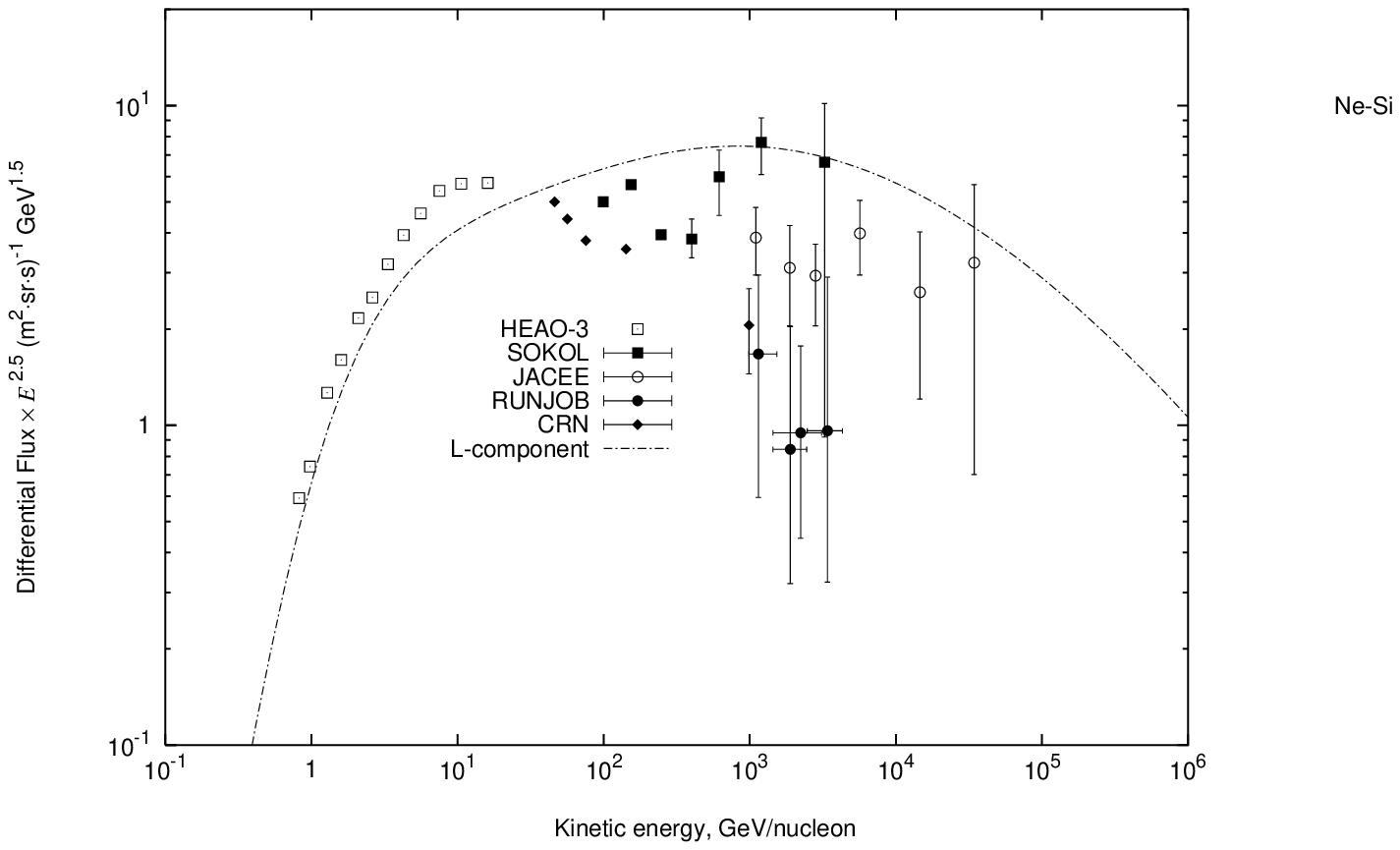}
\includegraphics[width=8.3cm]{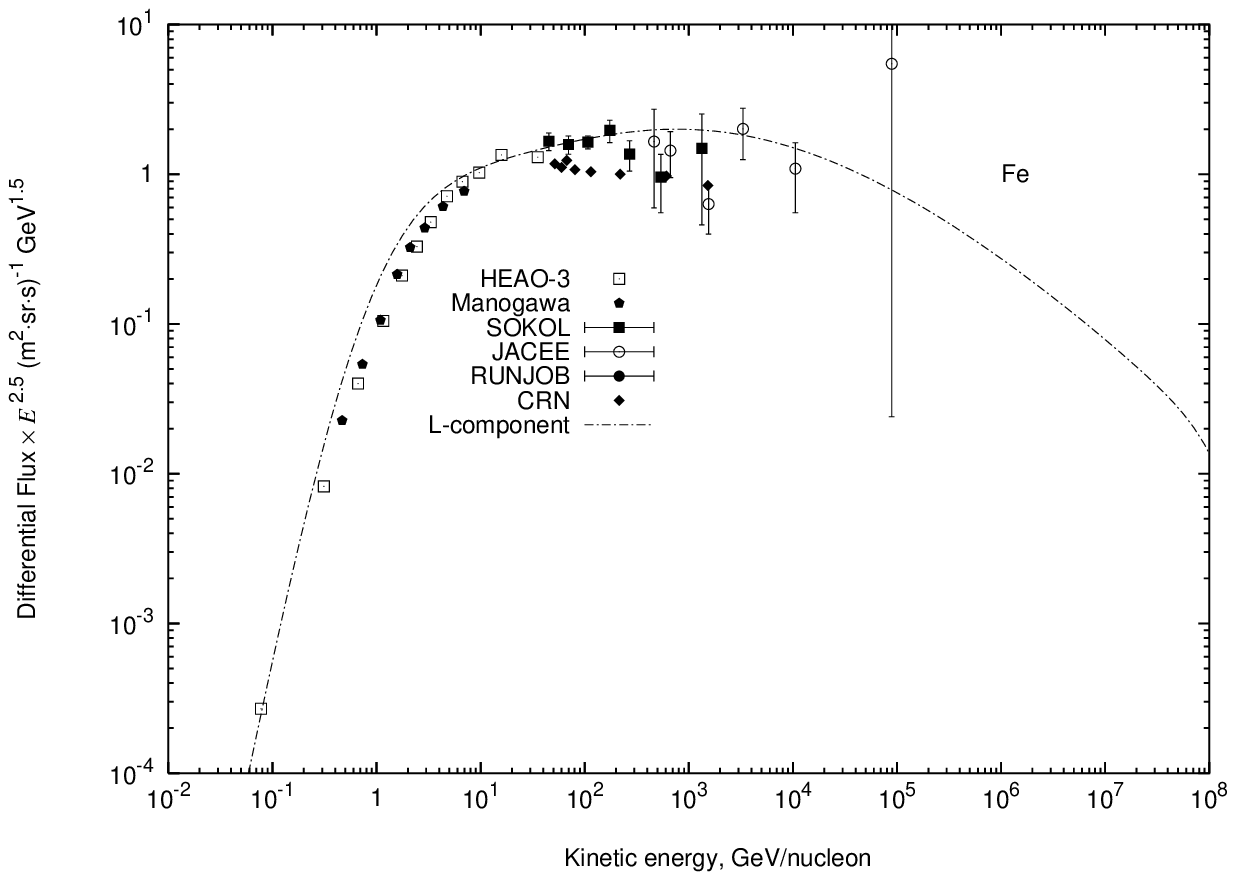}
\includegraphics[width=8.3cm]{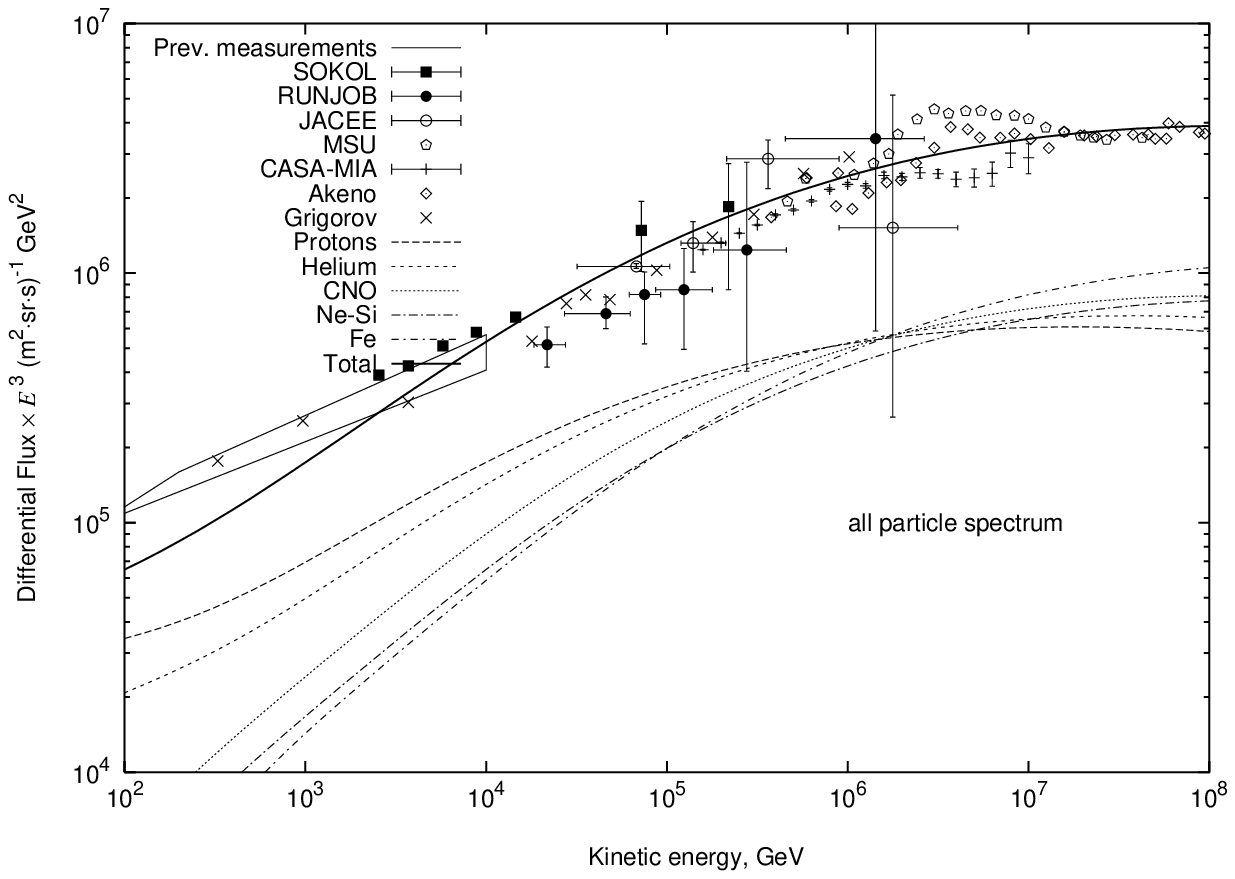}
\linebreak \caption{Comparison of our calculations of spectra with
experimental data:
\\ Grigirov --- \citet{Grigorov:1970},
\\ Ryan --- \citet{Ryan:1972},
\\ Minagawa --- \citet{Minagawa:1981},
\\ CRN --- \citet{Grunsfeld:1988},
\\ HEAO-3 --- \citet{Engel:1990},
\\ SOKOL --- \citet{Ivanenko:1990,Ivanenko:1993},
\\ MSU --- \citet{Fomin:1991},
\\ Swordy --- \citet{Muller:1991}
\\ Ichimura --- \citet{Ichimura:1993},
\\ Zatsepin --- \citet{Zatsepin:1993},
\\ Buckley --- \citet{Buckley:1994},
\\ Papiny, Lezniak --- \citet{Wiebel},
\\ JACEE --- \citet{Asakimory:1998},
\\ CAPRICE --- \citet{b108},
\\ CASA-MIA --- \citet{Glasmacher:1999},
\\ AKENO --- \citet{Yoshida:1995},
\\ RUNJOB --- \citet{b109}.}\label{fig1}
\end{figure*}
\section{Spectra and mass composition}
The differential flux $J_i$ of the particles of type $i$ due to
all sources of Galaxy may be separated into two components
\begin{equation}\label{J_i}
  J=J_L(r\leq 1 kpc)+J_G(r> 1 kpc).
\end{equation}
The first component (L) in (\ref{J_i}) describes the contribution
of the nearby sources (at distance $r\leq 1 kpc$) to observed flux
$J_i$. The second component (G) is the contribution of the distant
sources ($r> 1 kpc$) to $J_i$. The similar separation is
frequently used in the studies of cosmic rays (see, for instance,
\citet{Atoyan:1995}).

The list of nearby sources including 16 supernova remnants
\citep{Nichimura1, Nichimura2, Lozinskaya} is used to calculate
the L-component. The distant sources are supposed to be
distributed uniformly both in space and time. In this case
$J_G(r>1 kpc)\sim E^{-p-\delta/\beta}$ (see \citet{steady-state
sol.}).

Based on this result and (\ref{N(r,t,E}), we present the
differential flux in the form:
\begin{equation}\label{J result}
  J_i(E)=\frac{\upsilon_i}{4\pi}\sum_{\substack{i \\ (r\leq 1
  kpc)}}{N(\vec{r_i},t_i,E)+\upsilon_i C_{0i}
  E^{-p-\delta/\beta}}.
\end{equation}

It is clear from physical point of view that the bulk of observed
cosmic rays with energy $10^8\div 10^{10} eV$ forms by numerous
distant sources. It means that the observed flux at least of
protons and He in this energy region must be described by second
term in (\ref{J result}). The first term in (\ref{J result})
defines the spectrum in the high energy region and, as has been
shown above,  provides the ``knee''.

We use the spherically symmetric force model of \citet{Axford} to
describe the solar modulation. The influence of solar modulation
on the particle flux is
\[
J_{mod}(T)=\frac{T^2+2m_pc^2T}{(T+\Phi)^2+2m_pc^2(T+\Phi)}J_{ISM}(T+\Phi),
\]
where $T$ is the kinetic energy per nucleon, $m_p$ is the mass of
a proton and $J_{ISM}$ is the interstellar flux (\ref{J result}).
The potential energy $\Phi$, describing the average energy loss of
particle from interstellar space to 1 AU, is determined by solar
modulation parameter $\phi:\ \Phi=\phi Z/A$. $\phi=750 MV$ is
accepted in this paper (see \citet{b108,Menn:2000}).

The results of our calculation are presented in Fig. \ref{fig1},
and Table \ref{tab1}.
\begin{table}
\begin{tabular}{c|c|c|c|c|c|c|c} \hline
E,Gev&H&He&CNO&Ne-Si&Fe&$\langle\ln A\rangle$&$\langle A\rangle$\\
\hline 1E+2&0.53&0.32&0.08&0.05&0.02&0.90&5.36\\
3E+2&0.45&0.30&0.12&0.08&0.06&1.20&8.41\\
1E+3&0.40&0.29&0.14&0.10&0.08&1.40&10.38\\
3E+3&0.36&0.28&0.16&0.11&0.10&1.53&11.65\\
1E+4&0.33&0.27&0.17&0.12&0.11&1.65&12.91\\
3E+4&0.30&0.26&0.18&0.14&0.13&1.79&14.37\\
1E+5&0.26&0.24&0.19&0.15&0.15&1.93&15.95\\
3E+5&0.24&0.23&0.20&0.16&0.17&2.06&17.54\\
1E+6&0.21&0.21&0.20&0.17&0.20&2.17&19.06\\
3E+6&0.19&0.20&0.21&0.18&0.22&2.28&20.44\\
1E+7&0.18&0.19&0.21&0.19&0.24&2.37&21.68\\
3E+7&0.16&0.18&0.21&0.19&0.26&2.45&22.76\\
1E+8&0.15&0.17&0.21&0.20&0.27&2.51&23.70\\
1E+8&0.15&0.17&0.21&0.20&0.27&2.52&23.87\\ \hline
\end{tabular}
\caption{Mass composition of cosmic rays in anomalous diffusion
model }\label{tab1}
\end{table}
\balance
\section{Conclusion}
We have considered the propagation of galactic cosmic rays in the
fractal interstellar medium. The energy spectra of nuclei (H, He,
CNO, Ne-Si, Fe) and mass composition have been calculated in
energy region ($1\div 10^8$) GeV/particle under the natural
physical conditions $D(E)\sim E^\delta,\ S(E)\sim E^{-p}$. We have
shown that the model can explain the different values of spectral
exponent of protons and other nuclei, mass composition variations
at $E\sim 10^2\div 10^5$ GeV/nucleon, the steepening of the
all-particle spectrum.
\begin{acknowledgements}
This work was supported by the program ``Integration'' (project
2.1.--252).
\end{acknowledgements}

\end{document}